\documentclass[aps,superscriptaddress,twocolumn,showpacs,preprintnumbers,amsmath,amssymb]{revtex4}
\usepackage[cp1251]{inputenc}
\usepackage[english]{babel}
\usepackage{amssymb,amsmath}
\usepackage{amstext} 
\usepackage[dvips]{hyperref}
\usepackage[mathscr]{eucal}
\usepackage{longtable}
\usepackage{graphicx}
\begin{document}
\title{Rabi Waves  in Carbon Nanotubes}
\author{ Alla Dovlatova (a), Yauhen Yerchak (b) Dmitry Yearchuck (c)\\
\textit{(a) - M.V.Lomonosov Moscow State University, Moscow, 119899, \\ (b) - Belarusian State University, Nezavisimosti Ave., 4, Minsk, 220030, RB; \\ (c) - Minsk State Higher Aviation College, Uborevich Str., 77, Minsk, 220096, RB; yearchuck@gmail.com,}}
\date{\today}
\begin{abstract}QED-model for the multichain qubit system with interactions of qubits and chains between themselves on the example of the system of $\sigma$-polarons in carbon zigzag nanotubes, interacting with quantized EM-field, is considered analytically. The possibility of experimental detection of Rabi waves in conventional stationary optical experiments for any quasi-1D system with strong electron-photon interaction is predicted.\end{abstract}
\maketitle 
 Quantum electrodynamics (QED) consideration of interaction of electromagnetic (EM) field with matter is substantial for many practical applications, for instance  for
implementing of quantum computation, and it is studied extensively in two main directions. They are the Jaynes-Cummings  model (JCM) extending by considering of  multiple atoms
and multiple mode field and JCM-extending by introducing of
 nonlinearity in   JCM-Hamiltonian.   The examples of the synthesis of given two directions  are  recent works \cite{Slepyan}, \cite{Slepyan_Yerchak}, where Tavis-Cummings model \cite{Tavis} is developed by taking into account the interaction between qubits on the one hand  and simultaneously a nonlinearity is also included into consideration on the other hand. The authors  predict  the new coherent effect of nonlinear quantum optics -- spatial propagation of Rabi oscillations (Rabi waves) in 1D quantum dot  chain.
It represents the interest in a QED-consideration of the interaction with EM-field  of multichain systems by including the interchain interaction. It is aim of the paper presented. Especially interesting to consider practically significant case of carbon nanotubes (NT) to be a matter subsystem of the system matter + quantized EM-field. Really, for instance, NT of zigzag kind and NT of armchair kind  can be considered  to be the set of trans-polyacetylene (t-PA) and cis-polyacetylene (c-PA) chains correspondingly,  which are connected between themselves. There seems to be the most interesting  to consider  zigzag carbon NT, since the theory  of electronic properties of  t-PA chains is very well developed and it has very good experimental confirmation. Let us touch on given subject in more detail. Pioneering works \cite{Su_Schrieffer_Heeger_1979} \cite{Su_Schrieffer_Heeger_1980} were opening  new era in the physics of conjugated 1D conductors. The authors have found the most simple way to describe mathematically rather complicated system - the chain of t-PA, electronic system of which represents itself the example of Fermi liquid. It is now well known SSH-model. The most substantial suggestion in SSH-model is the suggestion, that the only dimerization coordinate $u_n$ of the $n$-th $CH$-group, 
$n = \overline{1,N}$ along chain molecular-symmetry axis $x$ is substantial for determination of main physical properties of the material. Other five degrees of freedom were not taken into consideration. Nevertheless, the model has obtained magnificent experimental confirmation. Naturally, it is reasonable to suggest, that given success is the consequence of some general princip.   Given general princip is really exists and main idea was proposed by Slater at the earliest stage of quantum physics era already in 1924. It is - "Any atom may in fact be supposed to communicate with other atoms all the time it is in stationary state, by means of virtual field of radiation, originating from oscillators having the frequencies of possible quantum transitions..." \cite{Slater}. Given idea obtains its development, if to clarify  the origin  of virtual field of radiation. It is shown in \cite {Dovlatova}, that Coulomb field in 1D systems has the character of radiation field and it can exist without the sources, which have created given field. It can be considered to be virtual field in Slater princip.  It produces preferential direction in atom  communication the only along chain axis, and it explains qualitatively the success of SSH-model. 
It is remarkable, that SSH-model contains in implicit form along with the physical basis of the existence  solitons, polarons, breathers, formed in $\pi$-electronic subsystem ($\pi$-solitons, $\pi$-polarons, $\pi$-breathers), also the basis for the existence of similar quasiparticles in  $\sigma$-electronic subsystem, that is $\sigma$-solitons, $\sigma$-polarons, $\sigma$-breathers. The cause  is the same two-fold degeneration of ground state  of the whole electronic system, energy of which has the form of Coleman-Weinberg potential with two minima at the values of dimerization coordinate $u_0$ and $ -u_0$.    SSH-$\sigma$-polarons have recently been experimentally detected in related material - carbynes \cite{Yearchuck_PL}, where the formation of polaron lattice  was
proposed.    Corresponding chain state is optically active and it is characterized by  the set of lines in IR-spectra, which were assigned with new phenomenon - antiferroelectric spin wave resonance (AFESWR). Slater princip allows to suggest, that the picture in zigzag NT can be quite similar, that is, despite the strong interaction between the chains, the physics of propagation processes will be very similar to that in t-PA and carbynes and propagation process in zigzag NT will represent the sum of $n$ individual processes.
We can now return to the problem put by. In the frames of SSH-model zigzag NT represents itself autonomous dynamical system with discrete circular symmetry consisting of finite number $n\in N$ of t-PA chains, which are placed periodically along transverse angle coordinate. Longitudinal axes $\{x_i\}, i = \overline{1,n}$, of individual chains can be directed both along element of cylinder and  along generatrix  of any other smooth figure with axial symmetry. It is taken into account, that in the frames of SSH-model, which take into consideration the dynamical processes the only along chain direction, the adjacent chains, which represent themselves a mirror of each other, will be indistinguishable. Here we will consider the only SSH-$\sigma$-polarons to be optically active centers, which produce polaron lattice. Each $\sigma$-polaron in accordance with experiment \cite{Yearchuck_PL} can be approximated like to guantum dot in \cite{Slepyan} by two-level gubit. Then for any individual chain the Hamiltonian, proposed in
\cite{Slepyan_Yerchak} can be used. Further to take into consideration all the rest chains and the interaction between them, we will use the apparatus of hypercomplex $n$-numbers. Let us remember briefly, that hypercomplex $n$-numbers are defined to be elements of commutative ring
\begin{equation}
\label{Eq5}
Z_n = C \oplus {C} \oplus{...}\oplus {C},
\end{equation}
that is, it is direct sum of $n$ fields of complex numbers $C$, $n\in N$. It means, that any hypercomplex $n$-number $z \in Z_n$ is $n$-dimensional quantity with the components $k_\alpha \in C, \alpha = \overline{0, n-1}$, and it in row matrix form $z \in Z_n$ is  
\begin{equation}
\label{Eq6}
z = [k_\alpha] = [k_0, k_1, ..., k_{n-1}].
\end{equation}
It can be represented in the form
\begin{equation}
\label{Eq7}
z = \sum_{\alpha = 0}^{n-1}k_\alpha\pi_\alpha,
\end{equation}
where $\pi_\alpha$ are basis elements of $Z_n$ (and simultaneously  basis elements of the linear space of n-dimensional lines and  n-dimensional row matrix). They are
\begin{equation}
\label{Eq8}
\begin{split}
&\pi_0 = [1,0, ...,0,0],  \pi_1 =[0,1, ...,0,0],\\
&..., \pi_{n-1} = [0,0, ...,0,1].
\end{split}
\end{equation}
Basis elements $\pi_\alpha$ possess by projection properties
\begin{equation}
\label{Eq9}
\pi_\alpha\pi_\alpha = \pi_\alpha\delta_{\alpha\beta}, \sum_{\alpha = 0}^{n-1}\pi_\alpha = 1, z \pi_\alpha = k_\alpha\pi_\alpha
\end{equation}
In other words the set of $k_\alpha \in C, \alpha = \overline{0, n-1}$ is the set of eigenvalues of hypercomplex $n$-number $z \in Z_n$, the set  of $\{\pi_\alpha\}$, $\alpha = \overline{0, n-1}$ is eigenbasis of $Z_n$-algebra.
Then the QED-Hamiltonian, considered to be hypercomplex operator n-number, for $\sigma$-polaron system of zigzag $NT$, consisting of $n$ t-PA chains and connected between themselves in that way, in order to produce rolled up graphene sheet, and interacting with EM-field in matrix representation is
\begin{equation} 
\label{Eq10}
[\hat{\mathcal{H}}] = [\hat{\mathcal{H}}_{\sigma}] + [\hat{\mathcal{H}}_F] + [\hat{\mathcal{H}}_{\sigma F}] + [\hat{\mathcal{H}}_{\sigma \sigma}] +  [\hat{\mathcal{H}}_{LF}].
\end{equation}
The rotating wave approximation and  the single-mode approximation of EM-field are used. The generalization for a multimode case thereafter being simple.
All the components in (\ref{Eq10}) are considered to be hypercomplex operator $n$-numbers and they are the following. $[\hat{\mathcal{H}}_{\sigma}]$ represents the operator of $\sigma$-polaron subsystem energy in the absence of interaction between $\sigma$-polaron  themselves and with EM-field. It is
\begin{equation} 
\label{Eq11}
[\hat{\mathcal{H}}_{\sigma}] = (\hbar \omega _0/2) \sum_{j = 0}^{n-1}\sum_m {\hat {\sigma}^z_{mj}}[e_1]^j,
\end{equation}
  where $\hat {\sigma}^z_{mj} = \left|a_{mj}\right\rangle  \left\langle a_{mj} \right|-\left|b_mj\right\rangle  \left\langle b_{mj} \right|$ is $z$-transition operator between the ground and excited states of $m$-th $\sigma$-polaron in $j$-th chain. The second term 
\begin{equation} 
\label{Eq12}
[\hat {\mathcal{H}}_F] = \hbar \omega \sum_{j = 0}^{n-1}\hat {a}^+\hat {a}[e_1]^j 
\end{equation}
is the Hamiltonian of the free electromagnetic field, which is represented in the form of 
hypercomplex operator $n$-number.
The component of the Hamiltonian (\ref{Eq10})
\begin{equation}
\label{Eq13}
[\hat {\mathcal{H}}_{\sigma F}] =\hbar g \sum_{j = 0}^{n-1}\sum\limits_m {(\hat {\sigma }_{mj}^+\hat {a}e^{ikma} + \hat {\sigma }_{mj}^-\hat {a}^+e^{-ikma})}[e_1]^j 
\end{equation}
describes the interaction of $\sigma$-polaron sybsystem with EM-field, where $g$  is the interaction constant.
The Hamiltonian
\begin{equation}
\begin{split}
\label{Eq14}
&[\hat{\mathcal{H}}_{\sigma\sigma}] =
-\hbar \sum_{l = 0}^{n-1}\sum_{j = 0}^{n-1}\xi^{(1)}_{|l-j|}[e_1]^l\sum\limits_m \left|a_{mj} \right\rangle \left\langle a_{m+1,j}\right|[e_1]^j \\
&-\hbar\sum_{l = 0}^{n-1}\sum_{j = 0}^{n-1}\xi^{(1)}_{|l-j|}[e_1]^l\sum\limits_m \left|a_{mj} \right\rangle \left\langle a_{m-1,j}  \right| [e_1]^j\\
&-\hbar\sum_{l=0}^{n-1}\sum_{j = 0}^{n-1}\xi^{(2)}_{|l-j|}[e_1]^l\sum\limits_m  \left| b_{mj} \right\rangle \left\langle b_{m+1,j}\right|[e_1]^j\\ &-\hbar\sum_{l=0}^{n-1} \sum_{j = 0}^{n-1}\xi^{(2)}_{|l-j|}[e_1]^l\sum\limits_m\left| b_{mj} \right\rangle \left\langle b_{p-1,j} \right|[e_1]^j,
\end{split}
\end{equation}
where  $\hbar\xi^{(1,2)}_{|l-j|}$ are the energies, characterizing intrachain $(l = j)$ and interchain $(l \neq j)$ polaron-polaron interaction for the excited ($\xi^{(1)}$) and ground ($\xi^{(2)}$) states of $j$-th chain, $\left| b_{mj} \right\rangle, \left|a_{mj}\right\rangle$ are ground and excited states correspondingly of $m$-th $\sigma$-polaron of $j$-th chain. Hamiltonian in the form like to (\ref{Eq14}) at $n=0$ is usually used for description of tunneling 
between the states with equal energies, in particular for tunneling between quantum dot states \cite{Tsukanov}, \cite{Slepyan}. Hamiltonian (\ref{Eq14}) at any $n$ describes actually the connection between pairs of the states, which satisfy the following condition - the first state  in any pair results from the second state (and vice versa)  by time reversal. It is known, that for given states Cooper effect takes place, which, in particular, plays main role in BCS-theory of superconductivity. Therefore,  the application of Hamiltonian like to (\ref{Eq14}) is possible for any pair of time reversal symmetric states with equal energy.  The component of the Hamiltonian $[\hat{\mathcal{H}}_{LF}]$ corresponds to the local-field effects
originated from the dipole-dipole electron-hole intrapolaron interaction and it is
\begin{equation}
\label{Eq15}
 [\hat{\mathcal{H}}_{LF}] = \frac{4\pi}{V}(\overrightarrow{P}\tilde{\overline{D}}
\overrightarrow{P})\sum_{j = 0}^{n-1}\sum\limits_{m}(\hat{\sigma}^{-}_{mj}\langle\hat{\sigma}^{+}_{mj}
\rangle + \hat{\sigma}^{+}_{mj}\langle\hat{\sigma}_{mj}^{-} \rangle)[e_1]^j,
\end{equation}
where $\overrightarrow{P}$ is dipole moment of $\sigma$-polaron,  $\tilde{\overline{D}}$ is  depolarization tensor. In (\ref{Eq11}) to (\ref{Eq15}) $[e_1]^j$ is j-th power of the circulant matrix $[e_1]$, which is
\begin{equation}
\label{Eq16}
[e_1]=\left[\begin{array} {*{20}c} 0&1&0& ...&0  \\ 0&0&1& ...&0 \\ &...& \\ 0&0& ... &0&1\\1&0&...&0&0 \end{array}\right]
\end{equation}
By the way, if one
omits last two terms in Hamiltonian (\ref{Eq10}), it goes into $n$-chain generalization of
well-known Tavis-Cummings Hamiltonian \cite{Tavis}.
\begin{figure}
\includegraphics[width=0.5\textwidth]{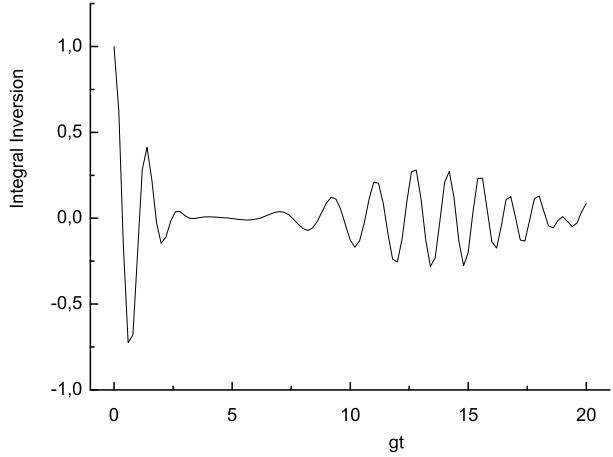}
\caption[Temporal dependence of the integral inversion  in the QD chain for a coherent initial state of light]
{\label{Figure1} Temporal dependence of the integral inversion  in the QD chain for a coherent initial state of light ($\left\langle n\right\rangle=4$).}
\end{figure}
The solution with given hypercomplex Hamiltonian and the solution for one chain in \cite{Slepyan_Yerchak} will formally have the same mathematical form.
\begin{figure}
\includegraphics[width=0.5\textwidth]{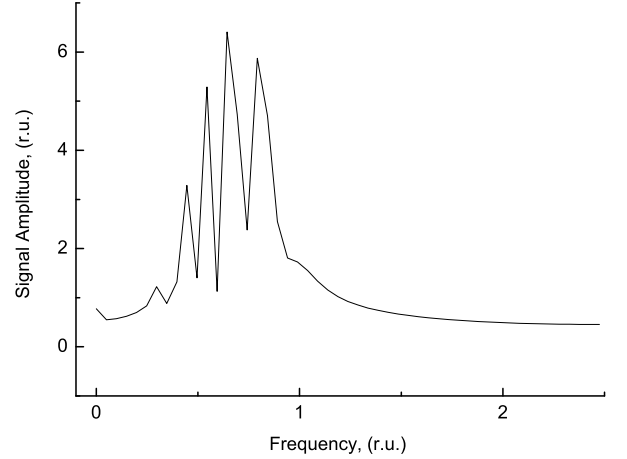}
\caption[Spectral dependence of the signal amplitude, corresponding to  temporal dependence, presented in Fig.1]
{\label{Figure2} Spectral dependence of the signal amplitude, corresponding to  temporal dependence, presented in Fig.1}
\end{figure}
 Following to \cite{Slepyan_Yerchak}, we will represent the state vector of the "NT+EM-field" system in terms of the eigenstates of isolated polaron and photon number states in the form
\begin{equation}
\begin{split}
\label{Eq17}
&[\left| {\Psi (t)} \right\rangle] = \\ 
&\sum_{j = 0}^{n-1}\{\sum\limits_l \sum\limits_{m} \left(A^j_{m,l}(t) \left|a_{mj},l \right\rangle + B^j_{m,l}(t) \left| b_{mj},l \right\rangle\right) \}[e_1]^j.
\end{split}
\end{equation}
Here, $\left| b_{mj},l \right\rangle = \left| b_{mj} \right\rangle\otimes\left|l \right\rangle$, $\left| a_{mj},l \right\rangle = \left| a_{mj} \right\rangle\otimes\left|l \right\rangle$, where  $\left|l \right\rangle$ is the EM-field  Fock state with $l$  photons, $A^j_{m,l}(t)$, $B^j_{m,l}(t)$ are the unknown probability amplitudes.
 Then, solving the nonstationary hypercomplex Schr\"odinger equation for the state vector $[\left| {\Psi (t)} \right\rangle]$, we obtain  in continuum limit
\begin{equation}
\begin{split}
\label{Eq18}
&[\Phi^l(x,t)] = \\
&\int\limits_{-\infty }^\infty{[\overline{\Phi}^l(h,0)]\exp\{i t([\theta^l(h)] - g \sqrt{l+1}[\chi])\}e^{ihx}}dh,
\end{split}
\end{equation}
where $x$ is  hypercomplex axis $x = [x, x, ..., x]$, $[\Phi^l(x,t)]$ is 
\begin{equation}
\begin{split}
\label{Eq19}
[\Phi^l(x,t)] = exp{\frac{i(\omega_0 t - kx)[\sigma_z]}{2}}\exp{\frac{\lambda t}{2}}[\Psi^l(x,t)], 
\end{split}
\end{equation}
in which $[\sigma_z]$ is Pauli z-matrix, $\lambda = \tau^{-1}$ is phenomenological relaxation factor. In its turn $[\Psi^l(x,t)]$ is continuous limit of functional block matrix of discrete variable $m$, which is
\begin{equation}
\begin{split}
\label{Eq20}
[\Psi_{m,l}(t)] = \left[\begin{array} {*{20}c}&[A_{m,l}(t)]\\&[B_{m,l+1}(t)]\end{array}\right],  
\end{split}
\end{equation}
that is, it is consisting of two $[n \times n]$ matrices of probability amplitudes
\begin{equation}
\begin{split}
\label{Eq21}
&[A_{m,l}(t)] = \sum_{j = 0}^{n-1}A^j_{m,l}(t)[e_1]^j, \\
&[B_{m,l+1}(t)] = \sum_{j = 0}^{n-1}B^j_{m,l+1}(t)[e_1]^j,
\end{split}
\end{equation}
which are determined by relationship (\ref{Eq17}).
 Further, matrix $[\theta(h)]$ in (\ref{Eq18}) is 
\begin{equation}
\begin{split}
\label{Eq22}
&[\theta(h)] = \frac{1}{2}\{([\theta_1(h)] + [\theta_2(h)]) \otimes [E_2]\} \\
&+ \frac{1}{2}\{([\theta_1(h)] - [\theta_2(h)]) \otimes [\sigma_z]\}, 
\end{split}
\end{equation}
where $[E_2]$ is unit $[2 \times 2]$-matrix, $[\theta_1(h)]$ and $[\theta_2(h)]$ are 
\begin{equation}
\begin{split}
\label{Eq23}
&[\theta_1(h)] = [\xi_1]\{2 - a^2(h + \frac{k}{2})^2\}, \\
&[\theta_2(h)] = [\xi_2]\{2 - a^2(h - \frac{k}{2})^2\}
\end{split}
\end{equation}
Here  $[\xi_1]$, $[\xi_2]$ are $[n \times n]$ matrices of coefficients in (\ref{Eq14}), that is
\begin{equation}
\begin{split}
\label{Eq24}
[\xi^j_1] = \sum_{l = 0}^{n-1}\xi^{(1)}_{|l-j|}[e_1]^l, [\xi^j_2] = \sum_{l = 0}^{n-1}\xi^{(2)}_{|l-j|}[e_1]^l,
\end{split}
\end{equation}
where one takes into account, that in view of axial symmetry $[\xi^j_{1,2}]$ do not depend on $j$. 
Consequently, we have $[\xi^j_1] = [\xi_1]$, $[\xi^j_2] = [\xi_2]$. Matrix $[\chi]$ in (\ref{Eq18}) is
\begin{equation}
\begin{split}
\label{Eq25}
[\chi] = [E_n] \otimes [\sigma_x]\exp {-i [\sigma_z] (\omega - \omega_0) t}.
\end{split}
\end{equation}
Matrix elements of $[\Phi^l(x,t)]$ are
\begin{equation}
\begin{split}
\label{Eq26}
&\Phi^l_{qp}(x,t) = \int\limits_{-\infty }^{\infty}\Theta^l_{q}(h,0) \exp{ \frac{-2\pi qpi}{n}} \exp{ihx}\times \\
&\exp{\{i\sum_{j = 0}^{n-1}\exp{\frac{2\pi qji}{n}(\vartheta_j(h) - g\sqrt{l-1}\kappa_j(h))}\}}dh,
\end{split}
\end{equation}
where $\Theta^l_{q}(h,0)$,$\vartheta_j(h)$, $\kappa_j(h)$ are determined by eigenvalues of $\Phi^l(h,0)$, $\theta(h)$ and $\chi(h)$, which are considered to be $n$-numbers. They are
\begin{equation}
\label{Eq27}
\Theta^l_{q}(h,0) = \frac{1}{n}\textbf{k}_q (\Phi^l(h,0)) = \frac{1}{n}\sum_{j = 0}^{n-1}\Phi_j^l(h,0)\exp{\frac{2\pi q j i}{n}}
\end{equation}
\begin{equation}
\label{Eq28}
\vartheta_j(h) = \frac{1}{n}\textbf{k}_j(\theta(h)) = \frac{1}{n}\sum_{r = 0}^{n-1}\theta_r(h)\exp{\frac{2\pi  jr i}{n}},
\end{equation}
\begin{equation}
\label{Eq29}
\kappa_j(h) = \frac{1}{n}\textbf{k}_j(\chi(h)) = \frac{1}{n}\sum_{r = 0}^{n-1}\chi_j(h)\exp{\frac{2\pi j r i}{n}}
\end{equation}
Owing to axial symmetry, the hypercomplex solution obtained can be represented in the form of sum of $n$ solutions for $n$ chains, that is hypercomplex $n$-number $\Phi^l(x,t) $ is
\begin{equation}
\label{Eq30}
\Phi^l(x,t) = \sum_{q = 0}^{n-1}\tilde{\Phi}^l_q(x,t),
\end{equation}
where the solution for $q$-th chain $\tilde{\Phi}^l_q(x,t)$ is
\begin{equation}
\label{Eq31}
\tilde{\Phi}^l_q(x,t) = \sum_{p = 0}^{n-1}\Phi^l_{qp}(x,t)[e_1]^p,
\end{equation}
in which the  matrix elements $\Phi^l_{qp}(x,t)$ are  determined by (\ref{Eq26}).
The relationship (\ref{Eq29}) by taking into account  (\ref{Eq18}) - (\ref{Eq28}) determines Rabi-wave packet, which propagates along individual chain of zigzag NT. It is evident, that the starting equation has also feasible solution in the form of traveling Rabi-waves. It is also clear, that analysis of Rabi-wave packet dynamics for individual NT-component will be the same, that in \cite{Slepyan_Yerchak}. Let us reproduce Fig.4 from \cite{Slepyan_Yerchak}, where temporal dependence of the integral inversion  in the QD chain for a coherent initial state of light ($\left\langle n\right\rangle=4$) is presented. It is seen, that an originally Gaussian packet temporally oscillates, at that oscillations collapse to zero quickly, but revive with time increasing in, and what is characteristic, in another area of space \cite{Slepyan_Yerchak}.  Temporal dependence (Fig.1) of the integral inversion gives in implicit form  the way for  comparison of theoretical results with any stationary optical experiments. 
Really, it is sufficient to make a Fourier transform of  given temporal dependence (Fig.2). It will be proportional to signal amplitudes of  infrared (IR) absorbance, IR-transmittance, IR-reflectance or Raman scattering, since they are  determined by population difference.
 Therefore, Rabi waves can be  registered  in 1D-systems by conventional IR- and RS-measurements, if the electron-photon coupling is rather strong. 
The comparison with experiment is given in \cite{D_Y}.

\end{document}